\documentclass[reprint,aps,pra,showpacs]{revtex4-1}

\usepackage{amsmath,graphicx,dcolumn,hyperref}
\usepackage{mathptmx,exscale}
\hypersetup{colorlinks=true,citecolor=blue,urlcolor=blue,linkcolor=blue}

\newcolumntype{d}[1]{D{.}{.}{#1}}

\renewcommand{\vec}{\mathbf}
\newcommand{\lmax}{l_\text{max}}
\newcommand{\nmax}{n_\text{max}}

\newcommand{\eps}{\varepsilon}

\newcommand{\tj}[6]{\begin{pmatrix} #1 & #2 & #3 \\ #4 & #5 & #6 \end{pmatrix}}
\newcommand{\sj}[6]{\begin{Bmatrix} #1 & #2 & #3 \\ #4 & #5 & #6 \end{Bmatrix}}

\let\originalleft\left
\let\originalright\right
\renewcommand{\left}{\mathopen{}\mathclose\bgroup\originalleft}
\renewcommand{\right}{\aftergroup\egroup\originalright}

\begin{document}
\frenchspacing

\title{Effective radius of ground- and excited-state positronium in collisions
with hard walls}
\author{R. Brown}
\altaffiliation[Present address: ]{School of Physics and Astronomy,
The University of Manchester, Manchester M13 9PL, United Kingdom}
\author{Q. Prigent}
\author{A. R. Swann}\email{aswann02@qub.ac.uk}
\author{G. F. Gribakin}\email{g.gribakin@qub.ac.uk}
\affiliation{School of Mathematics and Physics, Queen's University Belfast,
Belfast BT7 1NN, United Kingdom}
\date{\today}

\begin{abstract}
We determine effective collisional radii of positronium (Ps) by considering
Ps states in hard-wall spherical cavities. $B$-spline basis sets of
electron and positron states inside the cavity are used to construct the
states of Ps. Accurate Ps energy eigenvalues are obtained by extrapolation
with respect to the numbers of partial waves and radial states included in
the bases. Comparison of the extrapolated energies with those of a
pointlike particle provides values of the effective radius $\rho_{nl}$ of
Ps($nl$) in collisions with a hard wall. We show that for $1s$, $2s$, and
$2p$ states of Ps, the effective radius decreases with the increasing Ps
center-of-mass momentum, and find $\rho_{1s}=1.65$~a.u.,
$\rho_{2s}=7.00$~a.u., and $\rho_{2p}=5.35$~a.u. in the zero-momentum
limit.
\end{abstract}


\maketitle

\section{Introduction}

Positronium (Ps) is a light atom that consists of an electron and its
antiparticle, the positron. Positron- and positronium-annihilation-lifetime 
spectroscopy is a widely
used tool for studying materials, e.g., for determining pore sizes
and free volume. For smaller pores the radius of the Ps atom itself cannot be
neglected. This quantity was probed in a recent experiment which
measured the cavity shift of the Ps $1s$-$2p$ line~\cite{cassidylymanalpha},
and the data calls for proper theoretical understanding~\cite{greenreply}.
In this paper we calculate the eigenstates of Ps in a hard-wall spherical cavity
and determine the effective collisional radius of Ps in $1s$, $2s$, and $2p$
states as a function of its center-of-mass momentum.

The most common model for pore-size estimation is the
Tao-Eldrup model~\cite{tao,eldrup}. It considers an orthopositronium atom
($o$-Ps), i.e., a Ps atom in the triplet state, confined in the pore which is
assumed to be spherical, with radius $R_c$. The Ps is modeled as a point
particle with mass $2m_e$ in a spherical potential well, where $m_e$ is
the mass of an electron/positron. Collisions of $o$-Ps with the
cavity walls allow for positron two-gamma ($2\gamma$) annihilation with
the electrons in the wall, which reduces the $o$-Ps lifetime
with respect to the vacuum $3\gamma $-annihilation value of 142~ns. To simplify
the description of the penetration of the Ps wave function into the cavity wall,
the radius of the potential well is taken to be $R_c+\Delta R_c$, where the
best value of $\Delta R_c$ has been empirically determined to be
0.165~nm~\cite{schrader}. The model and its extensions are still widely used
for pore sizes in 1--100~nm range~\cite{gidley,goworek,wada}.

Porous materials and Ps confinement in cavities also enabled a number of
fundamental studies, such as measurement of Ps-Ps
interactions~\cite{cassidyinteractions}, detection of the Ps$_2$ molecule
\cite{cassidymolecule} and its optical spectroscopy~\cite{cassidyspectmol}, and
measurements of the cavity-induced shift of the Ps Lyman-$\alpha$ ($1s$-$2p$)
transition~\cite{cassidylymanalpha}. Cavities also hold prospects of
creating a Bose-Einstein condensate of Ps atoms and an
annihilation-gamma-ray laser~\cite{cassidymanypos}.

Seen in a wider context, the old subject of confined atoms~\cite{michels,sommerfeld} 
has seen renewed interest in recent years~\cite{jaskolski,buchachenko,connerade,atomicconfinement,sabin1,sabin2}. Studies
in this area not only serve as interesting thought experiments but also
apply to real physical situations, e.g., atoms under high pressure~\cite{lawrence,conneradecompress} or atoms trapped in fullerenes~\cite{bethune,shinohara,komatsu}. For $o$-Ps there is a specific question about
the extent to which confinement in a cavity affects its intrinsic $3\gamma $
annihilation rate (see Ref.~\cite{tanzi} and references therein).
 
For smaller cavities the effect of a finite radius of the trapped particle on
its center-of-mass motion cannot be ignored. In fact, the radius of a composite
quantum particle depends on the way this quantity is defined and probed. For
example, the proton is usually characterized by its root-mean-squared charge
radius. It is measured in elastic electron-proton scattering~\cite{bernauer} or
using spectroscopy of exotic atoms, such as muonic hydrogen~\cite{antognini}
(with as yet unexplained discrepancies between these experiments). For a particle
trapped in a cavity, any practically defined radius may depend on the nature of its
interaction with the walls.

In the present work we consider the simple problem of a Ps atom confined in a
hard-wall spherical cavity. The finite size of Ps gives rise to energy shifts
with respect to the energy levels of a pointlike particle in the cavity.
 This allows us
to calculate the effective \textit{collisional} radius of
Ps that describes its interaction with the impenetrable cavity wall. 

Ps is a hydrogenlike atom with a total mass of 2 and reduced mass of
$\frac12$ (in atomic units). The most probable distance between the electron
and positron in a free ground-state Ps atom is $2a_0$, where $a_0$ is the Bohr
radius, while the mean electron-positron separation is $3a_0$~\cite{landauQM}.
For excited states Ps$(nl$) these quantities increase as $n^2$. The Ps center
of mass is halfway between the two particles, so the most probable radius of
Ps($1s$) is $1a_0$, its mean radius being $1.5a_0$.  One can expect that the
distance of closest approach between the Ps center of mass and the wall with
which it collides will be similar to these values. One can also expect that
this distance will depend on the center-of-mass momentum of the Ps atom, as it
will be ``squashed'' when colliding with the wall at higher velocities.

A proper quantum-mechanical treatment of this problem is the subject of this
work. A configuration-interaction (CI) approach with a $B$-spline basis is
used to construct the states of Ps inside the cavity. Using these we determine the
dependence of the effective Ps radius on the center-of-mass momentum for the
$1s$, $2s$, and $2p$ states. 

Of course, the interaction between Ps and cavity walls in real materials
is different from the idealized situation considered here. It can be modeled
by changing the electron-wall and positron-wall potentials. On the other hand,
the hard-wall cavity can be used as a theoretical tool for studying Ps
interactions with atoms~\cite{psscattering}. An atom placed at the
center of the cavity will cause a shift of the Ps energy levels, whose positions
can be related to the Ps-atom scattering phase shifts $\delta_L(K)$ for the
$L$th partial wave~\cite{burke},
\begin{equation}\label{eqn:phaseshifts}
\tan\delta_L(K)=\frac{J_{L+1/2}(K [R_c-\rho(K)])}{Y_{L+1/2}(K [R_c-\rho(K)])},
\end{equation}
where $K$ is the Ps center-of-mass momentum, $J_\nu$ is the Bessel function,
$Y_\nu$ is the Neumann function, $R_c$ is the cavity radius, and $\rho(K)$ is the effective collisional radius of the Ps
atom.

The paper is organized as follows.
Section~\ref{sec:theory} describes the theory and numerical implementation of
the CI calculations of the energy levels and effective radii of Ps
in a spherical cavity. In Sec.~\ref{sec:results} these energies and radii are
presented for a number of cavity sizes and the dependence of the radii of
Ps($1s$), Ps($2s$), and Ps($2p$) on the Ps center-of-mass momentum is analyzed.
We conclude in Sec.~\ref{sec:conclusion} with a summary.

Unless otherwise stated, atomic units are used throughout.

\section{Theory and numerical implementation}\label{sec:theory}

\subsection{Ps states in the cavity}

The radial parts of the electron and positron states in an empty spherical
cavity with impenetrable walls are solutions of the Schr\"odinger equation
\begin{equation}
{-}\frac{1}{2}\frac{d^2P_{\eps l}}{dr^2}+\frac{l(l+1)}{2r^2}P_{\eps l}(r)=
\eps P_{\eps l}(r), \label{eq:Sch_rad}
\end{equation}
where $l$ is the orbital angular momentum, that satisfy the boundary conditions
$P_{\eps l}(0)=P_{\eps l}(R_c)=0$. Although Eq.~(\ref{eq:Sch_rad}) has analytical solutions,
we obtain the solutions numerically by expanding them in a $B$-spline basis,
\begin{equation}\label{eq:P_B} 
P_{\eps l}(r)=\sum _iC_iB_i(r),
\end{equation}
where $B_i(r)$ are the $B$ splines, defined on an equispaced knot sequence
\cite{cdeboor}. A set of 40 splines of order 6 has been used throughout.
Using $B$ splines has the advantange that a central atomic potential can be
added in Eq.~(\ref{eq:Sch_rad}) to investigate Ps-atom
interactions~\cite{psscattering}.

We denote the electron states by
$\phi_\mu(\vec{r}_e)=r_e^{-1}P_{\eps l}(r_e)Y_{lm}(\Omega _e)$, where
$Y_{lm}(\Omega )$ is the spherical harmonic that depends on the spherical angles
$\Omega $, and the positron states by $\phi_\nu(\vec{r}_p)$,
where $\vec{r}_e$ ($\vec{r}_p$) is the position vector of the electron
(positron) relative to the center of the cavity. For Ps in an empty cavity the
two sets of states are identical. The indices $\mu$ and $\nu$ stand for the
possible orbital angular momentum and radial quantum numbers of each state.

The nonrelativistic Hamiltonian  for Ps inside the cavity is
\begin{equation}
H = -\frac{1}{2} \nabla_e^2 -\frac{1}{2} \nabla_p^2 + V(\vec{r}_e,\vec{r}_p),
\end{equation}
where $V(\vec{r}_e,\vec{r}_p)=-|\vec{r}_e-\vec{r}_p|^{-1}$ is the Coulomb
interaction between the electron and positron. The infinite potential of the
wall is taken into account through the boundary conditions at $r_e=r_p=R_c$.
The Ps wave functions with a given total angular momentum $J$ and parity $\Pi$
are constructed as
\begin{equation}\label{eq:Pswfn}
\Psi_{J\Pi}(\vec{r}_e,\vec{r}_p) = \sum_{\mu ,\nu} C_{\mu\nu} \phi_\mu(\vec{r}_e)
\phi_\nu(\vec{r}_p),
\end{equation}
where the $C_{\mu\nu}$ are coefficients. The sum in Eq.~(\ref{eq:Pswfn}) is over
all allowed values of the orbital and radial quantum numbers up to infinity.
Numerical calculations employ finite values of $\lmax$ and $\nmax$,
respectively, and we use extrapolation to achieve completeness (see below).

Substitution of Eq.~(\ref{eq:Pswfn}) into the Schr\"odinger equation
\begin{equation}
H\Psi_{J\Pi}=E\Psi_{J\Pi},
\end{equation}
leads to a matrix-eigenvalue problem
\begin{equation}\label{eq:Hmat}
\vec{H} \vec{C}=E\vec{C},
\end{equation}
where the Hamiltonian matrix $\vec{H}$ has elements
\begin{equation}\label{eq:Hmael}
\langle \nu'\mu' \vert H \vert \mu\nu \rangle = ( \eps_\mu +
\eps_\nu ) \delta_{\mu\mu'}  \delta_{\nu\nu'} + \langle \nu'\mu' \vert V
\vert \mu\nu \rangle,
\end{equation}
$\eps_\mu$ ($\eps_\nu$) is the energy of the single-particle state $\mu$
($\nu$), and $\langle \nu'\mu' \vert V \vert \mu\nu \rangle$ is the
electron-positron Coulomb matrix element.
The vector $\vec{C}$ contains the expansion coefficients
$C_{\mu\nu}$. Diagonalization of the Hamiltonian matrix yields the energy
eigenvalues $E$ and the expansion coefficients. Working expressions for the
wave function and matrix elements, in which the radial and angular variables
are separated, are shown in Appendix~\ref{sec:app1}.

\subsection{Definition of the Ps effective radius}

The effective Ps radius is determined from the energy shifts with respect to the states of a point particle with the same mass as Ps.
We employ the notation $nl[N,L]$ to label the states of Ps in the cavity. Here
$nl$ refers to its internal state, and $[N,L]$ describes the state of the Ps
center-of-mass motion. The means of determining the four quantum numbers
$n$, $l$, $N$, and $L$ for Ps states is described in
Sec.~\ref{sec:statident}. Also, to obtain accurate values of the effective Ps
radius, the energy eigenvalues $E$ and other expectation values are
extrapolated to the limits $\lmax\to\infty$ and $\nmax\to\infty$; this is
discussed in detail in Sec.~\ref{sec:extrapolation}.

We consider each energy eigenvalue $E_{nl[N,L]}$ as the sum
\begin{equation}\label{eq:EnlNL}
E_{nl[N,L]}=E_{nl}^\text{int}+E_{nl[N,L]}^\text{COM},
\end{equation}
where $E_{nl}^\text{int}=-1/4n^2$ is the internal Ps bound-state energy, and
$E_{nl[N,L]}^\text{COM}$ is the energy of the center-of-mass motion.
The latter is related to the center-of-mass momentum
$K_{nl[N,L]}$ by
\begin{equation}\label{eq:ECOM}
E_{nl[N,L]}^\text{COM}=\frac{K_{nl[N,L]}^2}{2m},
\end{equation}
where $m=2$ is the mass of Ps.

Away from the wall the Ps wave function decouples into separate
internal and center-of-mass wave functions, viz.,
\begin{equation}\label{eq:Psi_nlNL}
\Psi_{J\Pi}(\vec{r},\vec{R}) \simeq \sum _{m,M_L}C^{JM}_{lmLM_L}
\psi_{nlm}^\text{int} (\vec{r}) \Phi_{nl[N,L]}^\text{COM}(\vec{R} ),
\end{equation}
where $\vec{r}=\vec{r}_e-\vec{r}_p$, $\vec{R}=(\vec{r}_e+\vec{r}_p)/2$ is
the position vector of the Ps center of mass, and $C^{JM}_{lmLM_L}$ is the
Clebsch-Gordan coefficient that couples the rotational state $lm$ of the Ps
internal motion with that of its center-of-mass motion ($LM_L$). Since the
center of mass is in free motion, its wave function is given by
\begin{equation}\label{eq:Phi_nlNL}
\Phi_{nl[N,L]}^\text{COM}(\vec{R}) \propto \frac{1}{\sqrt R}
J_{L+1/2}(K_{nl[N,L]}R) Y_{LM_L}(\Omega_\vec{R}).
\end{equation}
For a pointlike particle, the quantization of the radial motion in the
hard-wall cavity of radius $R_c$ gives $K_{nl[N,L]}R_c=z_{L+1/2,N}$, where
$z_{L+1/2,N}$ is the $N$th positive root of the Bessel function $J_{L+1/2}(z)$.
(For $S$-wave Ps, $L=0$, $z_{1/2,N}=\pi N$.) When the finite effective radius
$\rho_{nl[N,L]}$ of the Ps atom is taken into account, one has
\begin{equation}\label{eq:def_rho}
K_{nl[N,L]}(R_c-\rho_{nl[N,L]})=z_{L+1/2,N},
\end{equation}
which gives the energy (\ref{eq:EnlNL}) as
\begin{equation}\label{eq:EnlNLrho}
E_{nl[N,L]}=-\frac{1}{4n^2}+\frac{z_{L+1/2,N}^2}{4(R_c-\rho_{nl[N,L]})^2}.
\end{equation}
This relation defines the effective collisional radius of
Ps,
\begin{equation}\label{eq:rho_def}
\rho_{nl[N,L]} = R_c - z_{L+1/2,N}\left(4E_{nl[N,L]}+\frac{1}{n^2}\right)^{-1/2}.
\end{equation}
The Ps radius thus defined may depend on the Ps center-of-mass state $[N,L]$,
as well as the cavity radius. As we show in Sec.~\ref{sec:results}, the radius
is in fact determined only by the internal Ps state $nl$ and its center-of-mass
momentum $K$, i.e., it can be written as $\rho _{nl}(K)$.

In the present work, most of the calculations were performed for $J^\Pi=0^+$
and $1^-$, and cavity radii $R_c=10$~a.u. and 12~a.u. The value of $R_c$ was
kept small to assist convergence of the CI expansion (\ref{eq:Pswfn}).

\subsection{Identification of Ps states}\label{sec:statident}

After diagonalizing the Hamiltonian matrix and finding the energy
eigenvalues, one must determine the quantum numbers for each state before the
corresponding Ps radius $\rho_{nl[N,L]}$ can be calculated from
Eq.~(\ref{eq:rho_def}). To facilitate this, the mean electron-positron
separation $\langle r \rangle$ and contact density
$\langle \delta(\vec{r}) \rangle$ were calculated for each state (see
Appendix~\ref{sec:app1} for details).

The value of $\langle r \rangle$ for free Ps is twice the hydrogenic electron
radius~\cite{landauQM},
\begin{equation}\label{eq:meanr}
\langle r \rangle = 3n^2 - l(l+1).
\end{equation}
Thus, the expected mean separations for the $1s$, $2s$, and $2p$ states are 3, 12, and 10~a.u., respectively. In practice, the calculated
separations for the $2s$ and $2p$ states are noticeably lower (see Sec.~\ref{subsubsec:sep}), since the free-Ps values of $\langle r \rangle$ are comparable to the size of the cavity. Nevertheless, they are useful for identifying the individual Ps states.

The contact density $\langle \delta(\vec{r}) \rangle$ is useful for
distinguishing between $s$ and $p$ states, and between the $s$ states with
different principal quantum number $n$. For $s$ states of free Ps, the contact
density is given by the hydrogenic electron density at the origin
\cite{landauQM} scaled by the cube of the reduced-mass factor, viz.,
\begin{equation}
\langle \delta(\vec{r}) \rangle = \frac{1}{8\pi n^3}.
\end{equation}
Hence, the expected contact densities of $1s$ and $2s$ states are
$1/8\pi\approx 0.04$ and $1/64\pi \approx 5\times 10^{-3}$, respectively. For
$p$ and higher-angular-momentum states the contact density is zero. In
practice, the computed contact density for a $p$ state was observed to be of
the order of $10^{-8}$ or smaller.

Although looking at the numerical values of $\langle r \rangle$ and $\langle
\delta(\vec{r}) \rangle$ calculated for finite $\lmax$ and $\nmax$ is often
sufficient for distinguishing between the various states, their values can also
be extrapolated to the limits $\lmax\to\infty$ and $\nmax\to\infty$, as
demonstrated in Sec.~\ref{sec:extrapolation}.

Once the internal Ps state $nl$ has been established using the mean
electron-positron separation and contact density, the angular momentum
and parity selection rules,
\begin{gather}
\vert l-L \vert \leq J \leq l+L,\\
\Pi=(-1)^{l+L},
\end{gather}
allow one to determine the possible values of $L$. 

Finally, for fixed $n$, $l$, and $L$, with the energy eigenvalues arranged in increasing numerical order, the corresponding values of $N$ are 1, 2, 3, etc.

\subsection{Extrapolation}\label{sec:extrapolation}

\subsubsection{Energy eigenvalues}

In order to obtain the most precise values possible, the calculated energy
eigenvalues are extrapolated to the limits $\lmax\to\infty$ and
$\nmax\to\infty$. To this end, each calculation was performed for several
consecutive values of $\lmax$ with fixed $\nmax$, and this process was repeated
for several values of $\nmax$. For $J^\Pi=0^+$ we used values of $\lmax=11$--15
and $\nmax=10$--15. Due to computational restrictions~\footnote{We use an x86\textunderscore 64 Beowulf cluster.},
 for $J^\Pi=1^-$ it was
necessary to lower the values of $\lmax$ to the range 10--14, while keeping
$\nmax=10$--15 (Sec.~\ref{subsec:Jneq0}).
Extrapolation is first performed with respect to $\lmax$ for each value of
$\nmax$, and afterwards with respect to $\nmax$.

\begin{figure}[tb]
\centering
\includegraphics[width=\columnwidth]{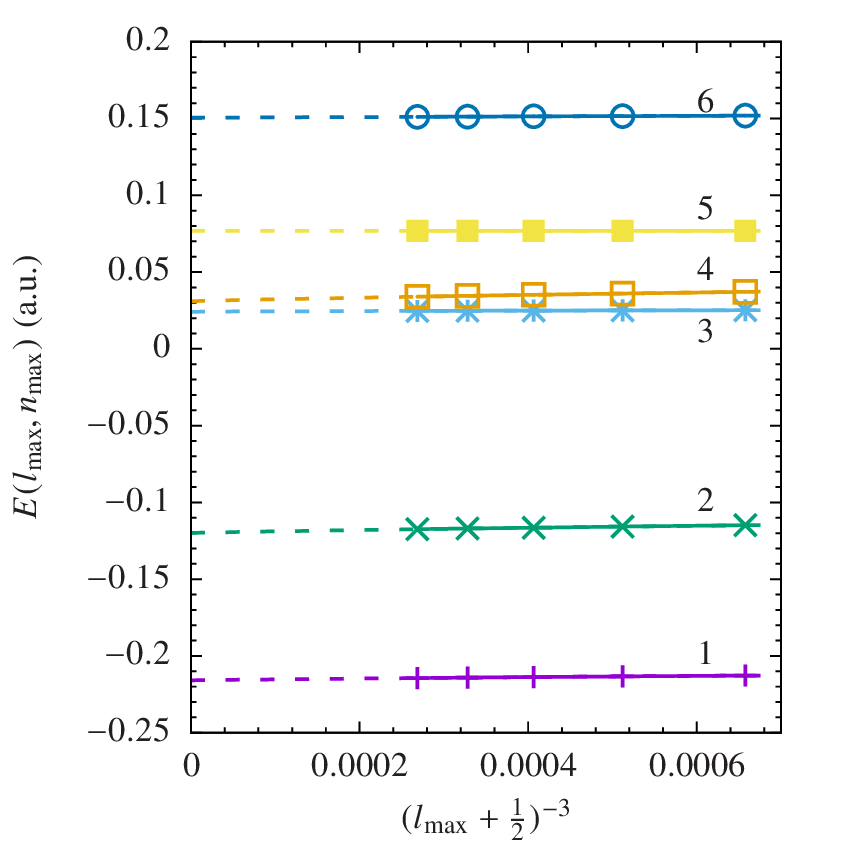}
\caption{Extrapolation in $\lmax$ for
the lowest 6 energy eigenvalues for $J^\Pi=0^+$, $R_c=10$~a.u., and
$\nmax=15$. Solid lines connect the data points to guide the eye, while the
dashed lines show extrapolation to $l_{\rm max}\rightarrow \infty$
using Eq.~(\ref{eq:Eextr_s}) or (\ref{eq:Eextr_p}).}
\label{fig:energyextraplmax}
\end{figure}

\begin{figure}[tb]
\centering
\includegraphics[width=\columnwidth]{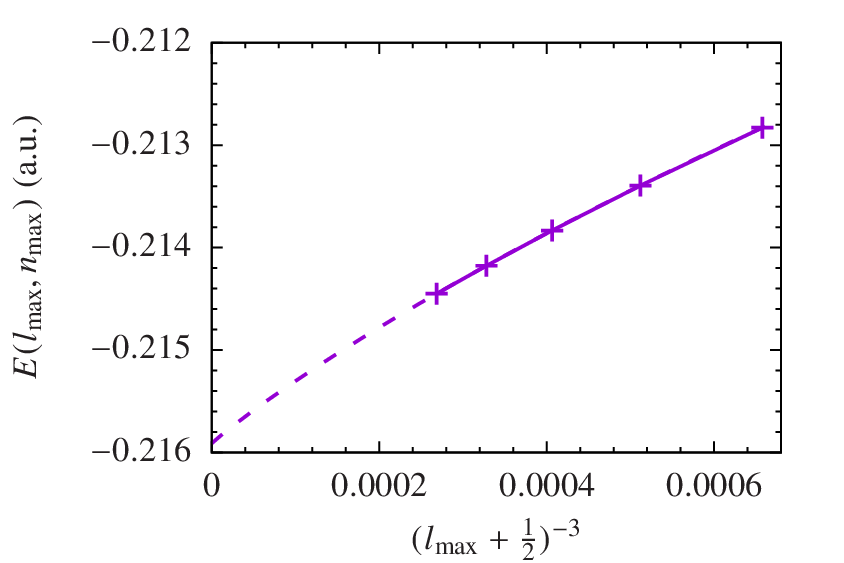}
\caption{Extrapolation in $\lmax$ for
the lowest energy eigenvalue for $J^\Pi=0^+$, $R_c=10$~a.u., and
$\nmax=15$. The solid line connects the data points to guide the eye, while the
dashed line shows extrapolation to $l_{\rm max}\rightarrow \infty$
using Eq.~(\ref{eq:Eextr_s}).}
\label{fig:en_extr_lmax_zoom}
\end{figure}

The convergence of CI expansions of the type (\ref{eq:Pswfn}) is controlled by
the Coulomb interaction between the particles. In our case, for $s$ states the
contributions to the total energy from electron and positron states with
orbital angular momentum $l$ behave as $(l+\frac12)^{-4}$,
while for $p$ states, as $(l+\frac12)^{-6}$~\cite{kutzelnigg,GL02}.
Denoting by $E(\lmax,\nmax)$ a generic unextrapolated energy eigenvalue
computed with partial waves up to $l_{\rm max}$ with $n_{\max }$ states in each,
we can extrapolate in $\lmax$ by using fitting curves of the form
\begin{align}\label{eq:Eextr_s}
E(\lmax,\nmax)&= E(\infty,\nmax)+A\left(\lmax+\tfrac12\right)^{-3}\nonumber\\
&\quad{}+ B\left(\lmax+\tfrac12\right)^{-4}+C\left(\lmax+\tfrac12\right)^{-5}
\end{align}
for $s$ states, and
\begin{align}\label{eq:Eextr_p}
E(\lmax,\nmax)&= E(\infty,\nmax)+A\left(\lmax+\tfrac12\right)^{-5}\nonumber\\
&\quad{}+ B\left(\lmax+\tfrac12\right)^{-6}
\end{align}
for $p$ states, where $A$, $B$, and $C$ are fitting parameters, and higher-order
terms are added to improve the fit. Such curves were found
to give excellent fits for our data points. Figure~\ref{fig:energyextraplmax}
shows the extrapolation of the 6 lowest energy eigenvalues for $J^\Pi=0^+$,
$R_c=10$~a.u., and $\nmax=15$. The effect of extrapolation can be seen better
in Fig.~\ref{fig:en_extr_lmax_zoom}, which shows it for the lowest eigenvalue.

\begin{figure}[tb]
\centering
\includegraphics[width=\columnwidth]{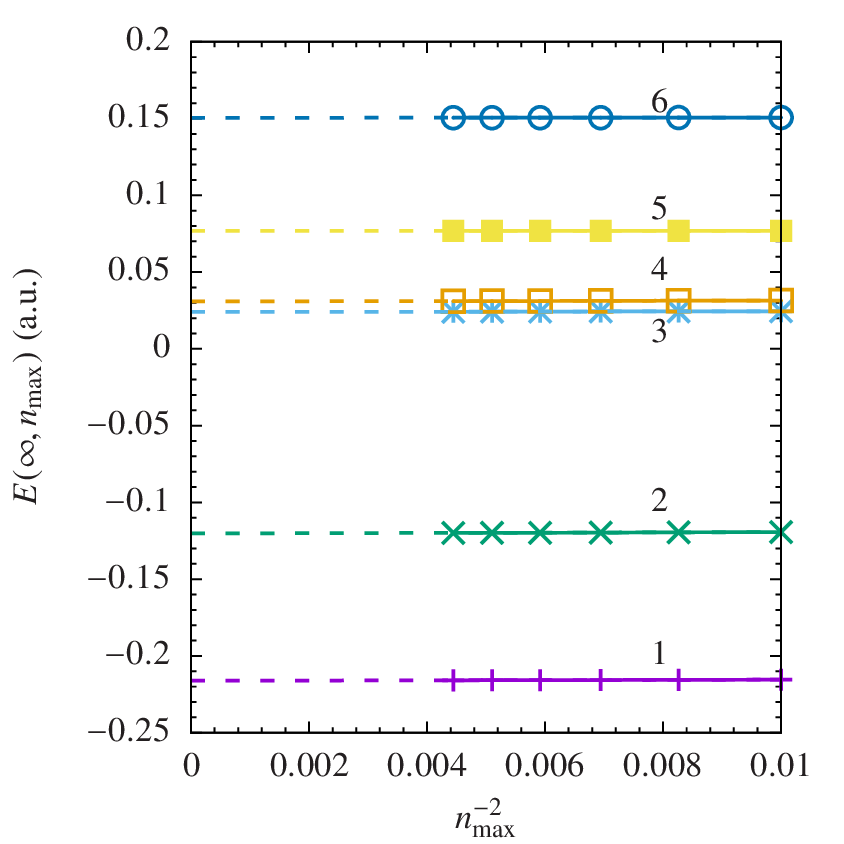}
\caption{\label{fig:energyextrapnmax}Extrapolation of the energy in $\nmax$ for
the lowest 6 energy eigenvalues for $J^\Pi=0^+$ and $R_c=10$~a.u. Solid lines
connect the data points to guide the eye, while the dashed lines show
extrapolation to $n_{\rm max}\rightarrow \infty$ using Eq.~(\ref{eq:Eextr_n}).}
\end{figure}

To extrapolate with respect to the maximum number of states in each
partial wave, we assume that the increments in the energy with increasing
$n_{\rm max}$ decrease as its negative power. The extrapolated energy eigenvalue
$E_{nl[N,L]}$ is then found from the fit
\begin{equation}\label{eq:Eextr_n}
E(\infty,\nmax)= E_{nl[N,L]} + \alpha\nmax^{-\beta},
\end{equation}
where $\alpha$ and $\beta$ are fitting parameters. Again, such curves produced
very good fits for our data points. Figure~\ref{fig:energyextrapnmax} shows the
extrapolation of the 6 lowest energy eigenvalues for $J^\Pi=0^+$ and
$R_c=10$~a.u.

It is known that CI-type or many-body-theory calculations for systems
containing Ps (either real or virtual) exhibit slow convergence with respect to
the number of partial waves included
\cite{bray,GL04,mitroybromley,zammit,Green13,green}.
Accordingly, the extrapolation in $\lmax$ is much more important than
the extrapolation in $\nmax$, which provides only a relatively small correction.
This can be seen in Table~\ref{tab:energies_radii}, which shows the final
extrapolated energy eigenvalues $E_{nl[N,L]}$ together with $E(\infty,\nmax)$
and $E(\lmax,\nmax)$ obtained for the largest $\lmax $ and $\nmax $ values.

\subsubsection{Electron-positron separation}\label{subsubsec:sep}

Expectation values of the electron-positron separation $\langle r \rangle$ and
contact
density $\langle \delta(\vec{r}) \rangle$ can also be extrapolated to the limits
$\lmax\to\infty$ and $\nmax\to\infty$. As with the energy, the extrapolation in
$\nmax$ makes only a small correction, which makes it superfluous here. We
only use these quantities to identify Ps states, so precise values are not needed.

\begin{figure}[tb]
\centering
\includegraphics[width=\columnwidth]{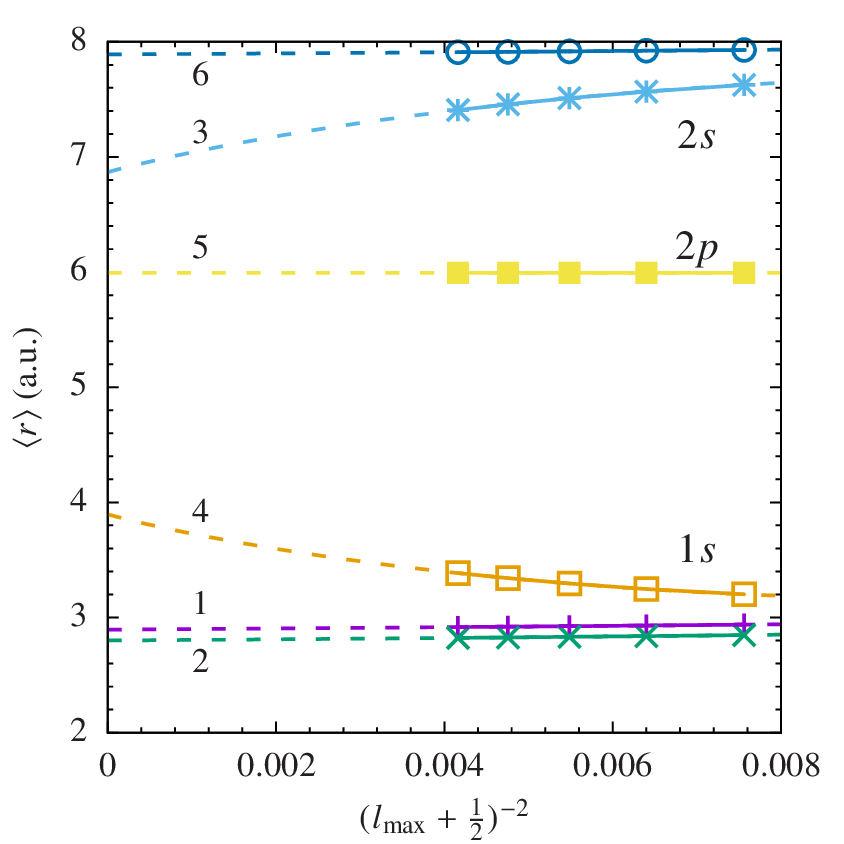}
\caption{Extrapolation of the expected
electron-positron separation in $\lmax$ for the lowest 6 energy eigenvalues
for $J^\Pi=0^+$, $R_c=10$~a.u., and $\nmax=15$. Solid lines connect the data
points to guide the eye, while the dashed lines show extrapolation to
$l_{\rm max}\rightarrow \infty$ using Eq.~(\ref{eq:sep_s_extrap}) or
(\ref{eq:sep_p_extrap}).}
\label{fig:separationextraplmax}
\end{figure}

For the mean separation we used fits of the form
\begin{equation}\label{eq:sep_s_extrap}
\langle r\rangle ^{[l_{\rm max}]} = \langle r\rangle + A \left( \lmax +
\tfrac12 \right)^{-2} + B \left( \lmax + \tfrac12 \right)^{-3}
\end{equation}
for $s$ states, and
\begin{equation}\label{eq:sep_p_extrap}
\langle r\rangle ^{[l_{\rm max}]} = \langle r\rangle + A \left( \lmax +
\tfrac12 \right)^{-4} + B \left( \lmax + \tfrac12
\right)^{-5}
\end{equation}
for $p$ states, where $\langle r\rangle ^{[l_{\rm max}]}$ is the value obtained
in the calculation with a given $l_{\rm max}$. Figure
\ref{fig:separationextraplmax} shows this extrapolation for $J^\Pi=0^+$,
$R_c=10$~a.u., and $\nmax=15$ for the 6 lowest energy eigenvalues, along with
tentative identifications of the quantum numbers $n$ and $l$.

The internal Ps states were determined as follows. States 1 and 2 appear to be
$1s$ states since $\langle r \rangle \approx 3$ for them. State 4 is also a
$1s$ state; its $\langle r \rangle$ values for smaller $\lmax $ are close to 3,
but increasing $\lmax $ and extrapolation lead to a higher value of
$\langle r \rangle \approx 4$. This distortion occurs due to level mixing
between the $1s$ state with $L=0$ and $N=3$ (state 4) and $2s$ state with
$L=0$ and $N=1$ (state 3, see below). These states are close in energy, and the
energy separation between them becomes smaller for
$l_{\rm max}\rightarrow \infty $ (see Fig.~\ref{fig:energyextraplmax}). This
increases the amount of level mixing and causes a noticeable decrease of the
expectation value of $\langle r \rangle $ with $l_{\rm max}$ for state 3. This
analysis is confirmed by the values of the contact density shown in
Sec.~\ref{subsubsec:contact}.

State 5 appears to be a $2p$ state due to the much larger value of
$\langle r \rangle$ compared with the $1s$ states, and also because an
excellent fit of the data points is obtained by using
Eq.~(\ref{eq:sep_p_extrap}), not Eq.~(\ref{eq:sep_s_extrap}). The value
of the contact density confirms this (Sec.~\ref{subsubsec:contact}). States 3
and 6 can be identified as $2s$ states because their mean separations are
higher than those of the $1s$ and $2p$ states [cf.~Eq.~(\ref{eq:meanr})], and
the data points are fitted correctly by using Eq.~(\ref{eq:sep_s_extrap}), not
Eq.~(\ref{eq:sep_p_extrap}). Note that the calculated separations for the
$2s$ and $2p$ states are lower than the free-Ps values of $12$~a.u. and
$10$~a.u. due to confinement by the cavity.

\subsubsection{Electron-positron contact density}\label{subsubsec:contact}

\begin{figure}[tb]
\centering
\includegraphics[width=\columnwidth]{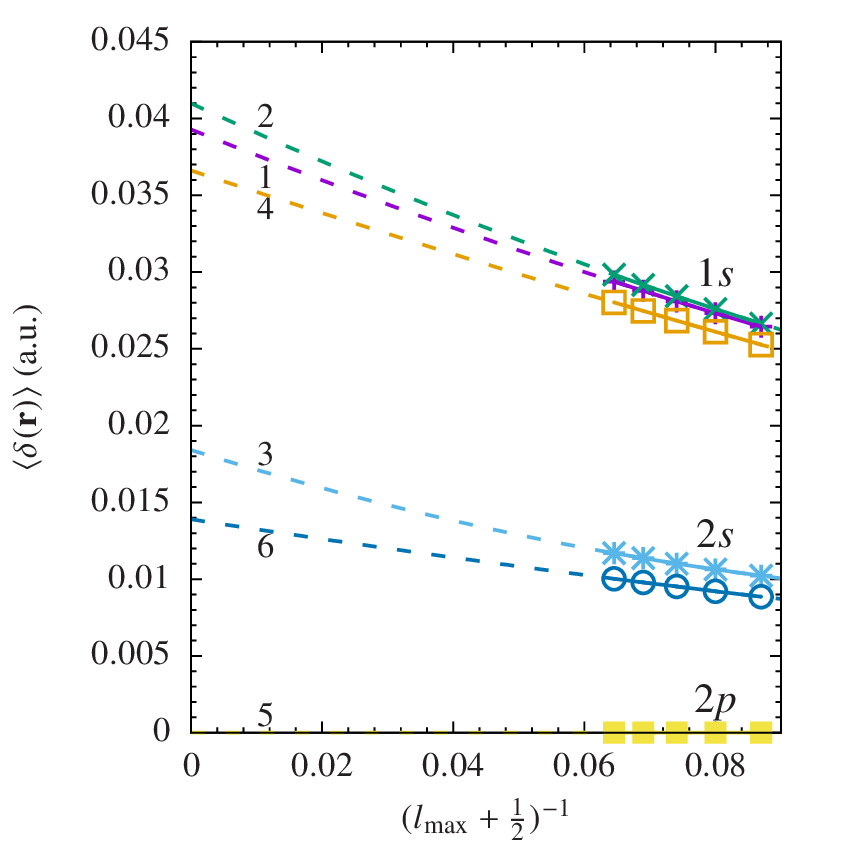}
\caption{Extrapolation of the expected contact
density in $\lmax$ for the 6 lowest-energy eigenstates for $J^\Pi=0^+$,
$R_c=10$~a.u., and $\nmax=15$. Solid lines connect the data
points to guide the eye, while the dashed lines show extrapolation to
$l_{\rm max}\rightarrow \infty$ using Eq.~(\ref{eq:contact_extrap}).}
\label{fig:contactextraplmax}
\end{figure}

Expectation values of the contact density $\langle \delta (\vec{r}) \rangle$
provide a useful check of the identification of the Ps states. This quantity
has the slowest rate of convergence in $\lmax$, and its extrapolation uses the
fit~\cite{GL02}
\begin{equation}\label{eq:contact_extrap}
\langle \delta (\vec{r}) \rangle ^{[l_{\rm max}]}=\langle \delta(\vec{r})\rangle
+ \frac{A}{\lmax + \tfrac12} + \frac{B}{\left( \lmax + \tfrac12 \right)^{2}}.
\end{equation}
Figure~\ref{fig:contactextraplmax} shows that for $R_c=10$~a.u. and $\lmax=15$,
extrapolation contributes up to 30\% of the final contact-density values
for the 6 lowest-energy $J^\Pi=0^+$ eigenstates.

Values of the contact density confirm the state identifications made in
Sec.~\ref{subsubsec:sep}. States 1, 2, and 4 have contact densities in the
range 0.037--0.041, close to the free-Ps value of $1/8\pi\approx0.0398$ for the
$1s$ state. State 5 has an extrapolated contact density of ${\sim} 10^{-15}$,
confirming that it is a $2p$ state (for which the free-Ps value is zero).

States 3 and 6, which we identify as $2s$ sates, have contact densities 0.018
and 0.014, respectively; the free-Ps value for the $2s$ state is
$1/64\pi \approx 0.005$, i.e., about 3 times smaller. The explanation for this
difference is that the confining cavity compresses the radial extent of the Ps
internal wave function, thereby increasing its density at $\vec{r}_e=\vec{r}_p$.
This effect has been observed in calculations of radially confined Ps
\cite{consolati}. The effect of compression on the contact density of Ps in
$2s$ states can be estimated from the ratio of the free-Ps mean distance
$\langle r\rangle =12$~a.u. to the values obtained in our calculation
(Fig.~\ref{fig:separationextraplmax}). The corresponding density enhancement is
proportional to the cube of this ratio, giving $(12/6.9)^3/64\pi \approx 0.026$
and $(12/7.9)^3/64\pi \approx 0.017$, for states 3 and 6, respectively, that
are close to the extrapolated densities in Fig.~\ref{fig:contactextraplmax}.
The same compression effect hardly affects $1s$ states because they are much
more compact, and the corresponding electron-positron separation values (for
states 1 and 2) are only a little smaller than the free-Ps value of 3~a.u.
As noted earlier, the $0^+$ states 3 ($1s$) and 4 ($2s$) exhibit some degree of
level mixing, which reduces the contact density of the former and increases that of
the latter.

\begin{table*}[tb]
\centering
\caption{Calculated energy eigenvalues, center-of-mass
momenta, and effective radii for $J^\Pi=0^+$, $1^-$ and $R_c=10$~a.u., 12~a.u.}
\label{tab:energies_radii}
\begin{ruledtabular}
\begin{tabular}{ccccD{.}{.}{1.8}D{.}{.}{1.8}D{.}{.}{1.8}D{.}{.}{1.6}D{.}{.}{1.3}}
$J^\Pi$ & $R_c$ & State no. & $nl[N,L]$ &
\multicolumn{1}{c}{$E(\lmax,\nmax)$\footnote{Energy eigenvalues obtained in the
largest calculation with $\nmax =15$ and $\lmax =15$ ($0^+$) or $\lmax =14$
($1^-$) for the positron.}} &
\multicolumn{1}{c}{$E(\infty ,\nmax)$\footnote{Energy eigenvalues obtained
after extrapolation in $\lmax $.}} &
\multicolumn{1}{c}{$E_{nl[N,L]}$\footnote{Energy eigenvalues obtained after
extrapolation in $\lmax $ and $\nmax$.}} &
\multicolumn{1}{c}{$K_{nl[N,L]}$} & \multicolumn{1}{r}{$\rho_{nl[N,L]}$} \\
\hline
$0^+$ & 10 & 1 & $1s[1,0]$ &-0.2144498&-0.215911&-0.216100& 0.368239 & 1.469 \\
&& 2 & $1s[2,0]$ & -0.1174946 & -0.119914 & -0.120167 & 0.720647 & 1.281 \\
&& 3 & $2s[1,0]$ & 0.02472915 & 0.0241658 & 0.0239796 & 0.588148 & 4.659 \\
&& 4 & $1s[3,0]$ & 0.03396135 & 0.0310792 & 0.0309400 & 1.060075 & 1.109 \\
&& 5 & $2p[1,1]$ & 0.07686479 & 0.0768633 & 0.0768631 & 0.746627 & 3.982 \\
&& 6 & $2s[2,0]$ & 0.1511621 & 0.150550 & 0.150504 & 0.923047 & 3.193 \\
\hline
$0^+$ & 12 & 1 & $1s[1,0]$ &-0.2251509&-0.227540&-0.227819 & 0.297866 & 1.453 \\
&& 2 & $1s[2,0]$ & -0.1593054 & -0.162912 & -0.163350 & 0.588727 & 1.328 \\
&& 3 & $1s[3,0]$ & -0.05568137 & -0.0601980 & -0.0605930 & 0.870418 & 1.172 \\
&& 4 & $2s[1,0]$ & -0.01046464 & -0.0107571 & -0.0108531 & 0.454519 & 5.088 \\
&& 5 & $2p[1,1]$ & 0.02539311 & 0.0253906 & 0.0253903 & 0.592926 & 4.422 \\
&& 6 & $2s[2,0]$ & 0.07401803 & 0.0721989 & 0.0719415 & 0.733325 & 3.432 \\
&& 7 & $1s[4,0]$ & 0.08558862 & 0.0811920 & 0.0809612 & 1.150585 & 1.078 \\
\hline
$1^-$ & 10 & 1 & $1s[1,1]$ &-0.1787461&-0.181426&-0.181500 & 0.523450 & 1.416 \\
&& 2 & $1s[2,1]$ &-0.05370988 & -0.0571538 & -0.0573473 & 0.877844 & 1.200 \\
&& 3 & $2p[1,0]$ & 0.004981346 & 0.00497852 & 0.00497844 & 0.519532 & 3.953 \\
&& 4 & $2s[1,1]$ & 0.08365396 & 0.0831681 & 0.0830909 & 0.763128 & 4.112 \\
&& 5 & $2p[1,2]$ & 0.1112176 & 0.111200 & 0.111199 & 0.833544 & 3.086 \\
&& 6 & $1s[3,1]$ & 0.1228439 & 0.118529 & 0.118321 & 1.213789 & 1.016 \\
&& 7 & $2p[2,0]$ & 0.1344290 & 0.134426 & 0.134425 & 0.887525 & 2.921 \\
\hline
$1^-$ & 12 & 1 & $1s[1,1]$ &-0.2004982&-0.204725&-0.204872& 0.424867 & 1.424 \\
&& 2 & $1s[2,1]$ & -0.1152794 & -0.120225 & -0.120512 & 0.719689 & 1.266 \\
&& 3 & $2p[1,0]$ & -0.02131484 & -0.0213199 & -0.0213201 & 0.405857 & 4.259 \\
&& 4 & $1s[3,1]$ & 0.006377168 & 6.09126\times10^{-4} & 1.97942\times10^{-4} &
1.000396 & 1.100 \\
&& 5 & $2s[1,1]$ & 0.02855653 & 0.0278117 & 0.0277551 & 0.600850 & 4.522 \\
&& 6 & $2p[1,2]$ & 0.05077695 & 0.0507504 & 0.0507503 & 0.673054 & 3.437 \\
&& 7 & $2p[2,0]$ & 0.06707912 & 0.0670723 & 0.0670721 & 0.719922 & 3.272
\end{tabular}
\end{ruledtabular}
\end{table*}

\subsection{Eigenstates with $J\neq 0$}\label{subsec:Jneq0}

Figures~\ref{fig:energyextraplmax}--\ref{fig:contactextraplmax} showed how the
accurate energies, electron-positron separations and contact densities of the 6
lowest-energy  $J^\Pi =0^+$ Ps eigenstates in the cavity of radius $R_c$ were
obtained. For this symmetry, the electron and positron orbital angular momenta
$l_\nu $ and $l_\mu $ in the expansion (\ref{eq:Pswfn}) are equal, and the
dimension of the Hamiltonian matrix in Eq.~(\ref{eq:Hmat}) is
$(\lmax +1)\nmax ^2$ (3600 in the largest
calculation). The ground state of the system describes Ps($1s$) with the
orbital angular momentum $L=0$ in the lowest state of the center-of-mass motion,
$N=1$. Higher-lying states correspond to excitations of the center-of-mass
motion of Ps($1s$) ($N>1$), as well as internal excitations of the Ps atom ($2s$
and $2p$). For $0^+$ symmetry, the center-of-mass orbital angular momentum
of Ps($2p$) is $L=1$, which is why this state (5 in
Fig.~\ref{fig:energyextraplmax}) lies higher than the lowest $L=0$, $N=1$ state
of Ps($2s$) (state 4).

We also calculated the eigenstates for a larger cavity radius $R_c=12$~a.u.
Increasing $R_c$ lowers the energies of all states, and for $J^\Pi =0^+$ we
identify four $1s$ states ($L=0$, $N=1$--4), two $2s$ states ($L=0$, $N=1$, 2),
and one $2p$ state ($L=1$, $N=1$); see Table~\ref{tab:energies_radii}.
States that lie at higher energies, above the Ps breakup threshold [$E=0$ for
free Ps, or above $2\pi ^2/(2R_c^2)\sim 0.1$~a.u. in the cavity] do not
have the form (\ref{eq:Psi_nlNL}) but describe a relatively weakly correlated
electron and positron ``bouncing'' inside the cavity.

To find other states of Ps($2p$) we performed calculations of $J^\Pi =1^-$
states for both $R_c=10$ and 12~a.u. For this symmetry the electron and
positron orbital angular momenta are related by $l_\mu =l_\nu \pm 1$, and the
size of the Hamiltonian matrix is $2\lmax \nmax ^2$, i.e., about a factor of
2 larger than for $J^\Pi =0^+$. For computational reasons, it is convenient to
define $\lmax $ as the maximum angular momentum of one of the particles, e.g.,
the positron. In this case the electron orbital angular momentum can be as
large as $\lmax +1$. Limiting its value by 14, we restrict
the value of $\lmax $ used for extrapolation to the range 10--13. This
difference aside, extrapolation of the energy eigenvalues and other quantities
for the $1^-$ states is performed as described in Sec.~\ref{sec:extrapolation}.
In total, we find 7 eigenstates for $J^\Pi=1^-$: 3 for Ps($1s$) ($L=0$,
$N=1$--3), 3 for Ps($2p$) ($L=0$, $N=1$, 2 and $L=2$, $N=1$), and 1 for Ps($2s$)
($L=1$, $N=1$); see Table~\ref{tab:energies_radii}.

\section{Results}\label{sec:results}

Table~\ref{tab:energies_radii} shows the quantum numbers and energy eigenvalues
$E_{nl[N,L]}$ of the $J^\Pi=0^+$ and $1^-$ states we found for $R_c=10$ and
12~a.u. alongside the corresponding Ps center-of-mass momenta $K_{nl[N,L]}$ and
effective Ps radii $\rho_{nl[N,L]}$. As expected, the values of $\rho_{nl[N,L]}$
for the Ps($1s$) states are much smaller than those for Ps in the
$2s$ and $2p$ states. The Ps radius for each internal state also displays
significant variation with the Ps center-of-mass quantum numbers $N$ and $L$
and with the cavity radius $R_c$. It turns out that, to a good approximation,
this variation can be analyzed in terms of a single parameter, namely, the
Ps center-of-mass momentum $K$.

\begin{figure}[tb]
\centering
\includegraphics[width=\columnwidth]{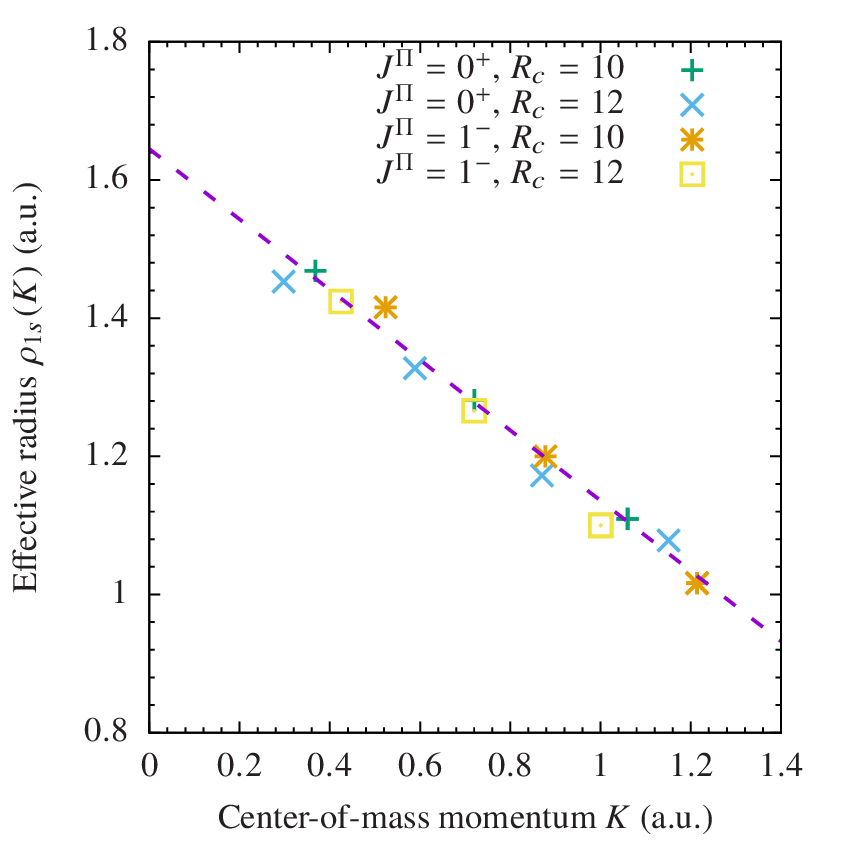}
\caption{Dependence of the effective Ps($1s$) radius
$\rho_{1s}(K)$ on the Ps center-of-mass momentum $K$. The dashed line is the
linear fit, Eq.~(\ref{eq:rho1s}).}
\label{fig:Ps1s}
\end{figure}

Figure~\ref{fig:Ps1s} presents 13 values of the radius of Ps($1s$) states from
Table~\ref{tab:energies_radii}, plotted as a function of $K$. 
The figure shows
that, to a very good approximation, the dependence of the Ps radius on
its center-of-mass momentum is linear, as described by the fit
\begin{equation}\label{eq:rho1s}
\rho_{1s}(K) = 1.65 - 0.51K.
\end{equation}
When Ps collides with the wall at higher momenta, its effective radius is
smaller. This could be expected from a naive picture of Ps as a ``soft ball''
that gets ``squashed'' when it hits the hard wall. For higher impact
velocities the distortion is stronger, and the Ps center of mass gets
closer to the wall. The predicted ``static'' (i.e., zero-incident-momentum)
radius for the $1s$ state is $\rho_{1s}(0)=1.65$~a.u. This value is close
to the mean distance between the Ps center of mass and either of its constituent
particles, $\frac12\langle r\rangle =1.5$~a.u.

\begin{figure}[tb]
\centering
\includegraphics[width=\columnwidth]{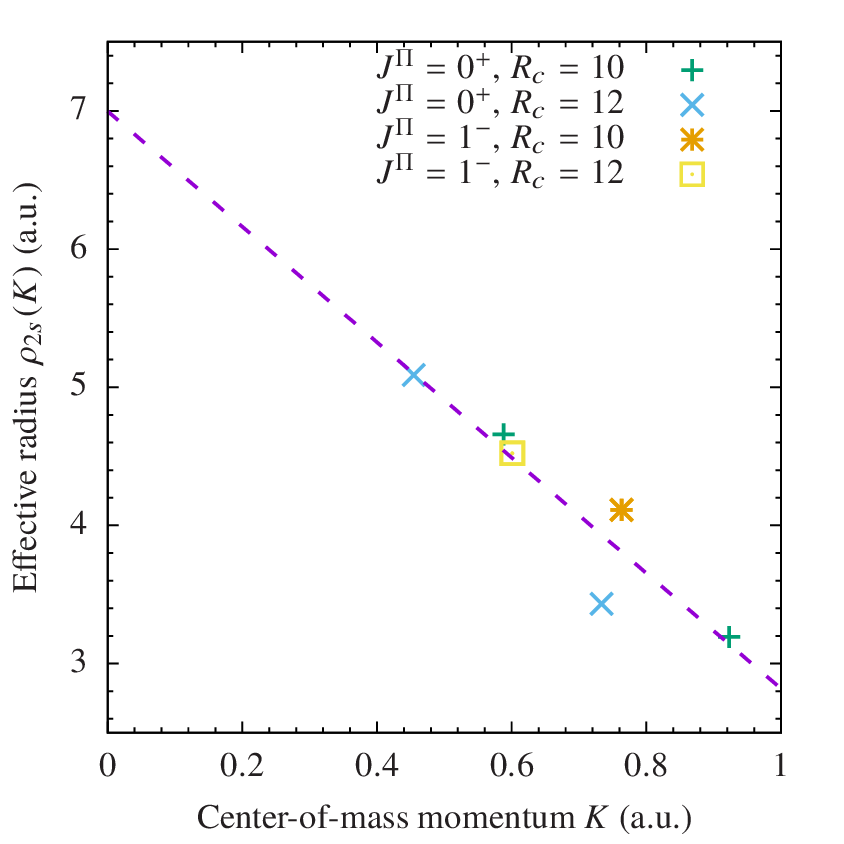}
\caption{\label{fig:Ps2s}Dependence of the effective Ps($2s$) radius
$\rho_{2s}(K)$ on the Ps center-of-mass momentum $K$. The dashed line is the
linear fit, Eq.~(\ref{eq:rho2s}).}
\end{figure}

Figure~\ref{fig:Ps2s} shows the data for the radius of the $2s$ state. 
They also suggest a near-linear relationship between the Ps radius and
center-of-mass momentum, with two points near $K\approx 0.75$ deviating
slightly from it. The larger deviation, observed for the $J^\Pi=0^+$,
$R_c=12$~a.u. datum (blue cross), could be, at least in part, due to level mixing
between eigenstates 6 and 7, which are separated by a small energy interval
(see Table~\ref{tab:energies_radii}). A linear fit of the data gives
\begin{equation}\label{eq:rho2s}
\rho_{2s}(K) = 7.00 - 4.18K.
\end{equation}
The corresponding static radius $\rho_{2s}(0)=7.00$~a.u. is close to
the mean radius of free Ps($2s$), $\frac12\langle r\rangle =6$~a.u.

\begin{figure}[tb]
\centering
\includegraphics[width=\columnwidth]{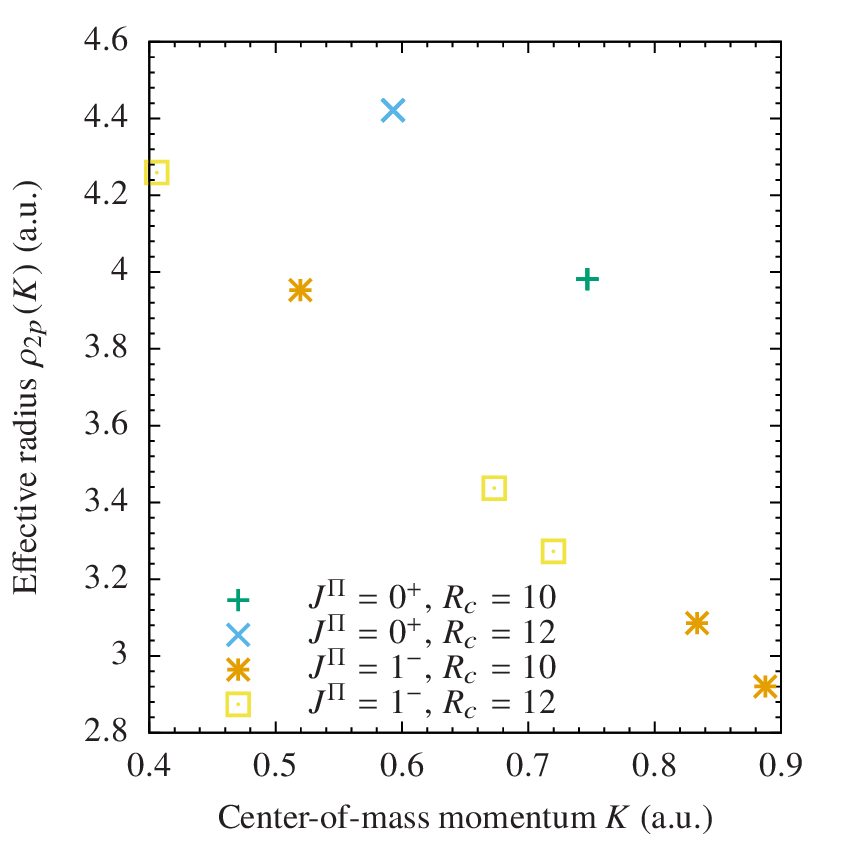}
\caption{Dependence of the effective Ps($2p$) radius
$\rho_{2p}(K)$ on the Ps center-of-mass momentum $K$.}
\label{fig:Ps2p}
\end{figure}

Finally, Fig.~\ref{fig:Ps2p} shows the dependence of the radius on the
center-of-mass momentum for the Ps($2p$) states. In this case, the 6 data
points for the $J^\Pi=1^-$ states again indicate a linear dependence of
$\rho_{2p}$ on $K$, while the points with $J^\Pi=0^+$ appear as outliers.
To understand this behavior, we performed additional sets of calculations
for $J^\Pi=1^+$ and $2^+$, with $R_c=10$ and 12~a.u. For $J^\Pi=1^+$ we used
$\lmax=10$--14 and $\nmax=10$--15, while for $J^\Pi=2^+$ we used $\lmax=9$--13
and $\nmax=8$--13, due to computational restrictions. For both values of $R_c$
we found one Ps($2p$) state with $N=1$ and $L=1$ for each symmetry.
Table~\ref{tab:1+2+} shows the data for these states.

\begin{table}[tb]
\caption{Calculated energy eigenvalues, center-of-mass momenta,
and effective radii for $J^\Pi=1^+$, $2^+$ and $R_c=10$ and 12~a.u.
Only the Ps($2p$) states are shown.}
\label{tab:1+2+}
\begin{ruledtabular}
\begin{tabular}{ccccD{.}{.}{1.7}D{.}{.}{1.6}D{.}{.}{1.3}}
$J^\Pi$ & $R_c$ & State no. & $[N,L]$ & \multicolumn{1}{c}{$E_{2p[N,L]}$} &
\multicolumn{1}{c}{$K_{2p[N,L]}$} & \multicolumn{1}{r}{$\rho_{2p[N,L]}$} \\
\hline
$1^+$ & 10 & 1 & $[1,1]$ & 0.0477391 & 0.664045 & 3.233 \\
$1^+$ & 12 & 1 & $[1,1]$ & 0.00718021 & 0.527940 & 3.489 \\
$2^+$ & 10 & 3 & $[1,1]$ & 0.0576334 & 0.693205 & 3.518 \\
$2^+$ & 12 & 3 & $[1,1]$ & 0.0134862 & 0.551312 & 3.850
\end{tabular}
\end{ruledtabular}
\end{table}

\begin{figure}[tb]
\centering
\includegraphics[width=\columnwidth]{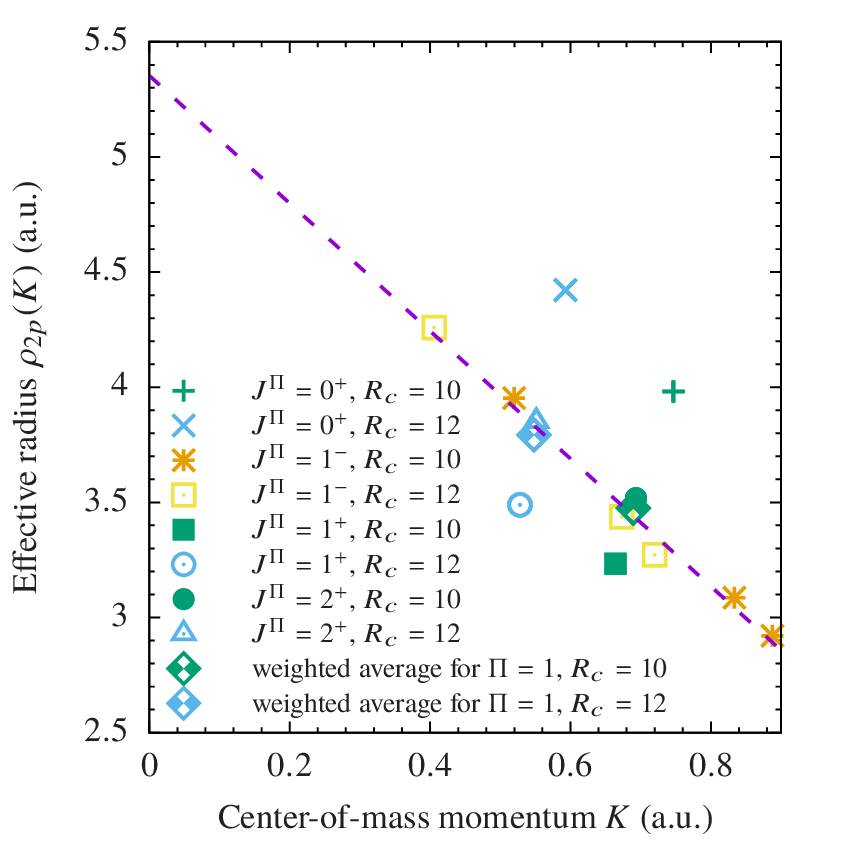}
\caption{Dependence of the effective Ps($2p$) radius $\rho_{2p}(K)$ on the Ps
center-of-mass momentum $K$, including data for $J^\Pi=1^+$ and $2^+$ states.
The dashed line is the linear fit, Eq.~(\ref{eq:rho2p}).}
\label{fig:Ps2p1+2+}
\end{figure}

Figure~\ref{fig:Ps2p1+2+} shows the dependence of the radius on the momentum for
the Ps($2p$) states, using the data from all of the calculations performed. 
As noted above, the negative-parity $J^\Pi=1^-$ data that describe the states
of Ps($2p$) with the center-of-mass angular momentum $L=0$ or 2 display a
clear linear trend. In contrast, the three positive-parity states with
$J^\Pi =0^+$, $1^+$, and $2^+$ (for a given $R_c$) do not follow the trend
and suggest $J$-dependent values of the Ps radius. These states
correspond to three possible ways of coupling the Ps($2p$) internal
angular momentum $l=1$ and its center-of-mass angular momentum $L=1$.
Since neither Ps($2p$) nor its center-of-mass wave function for $L=1$
[cf. Eq.~(\ref{eq:Psi_nlNL})] is spherically symmetric, it is not 
surprising that the distance of closest approach to the wall (i.e., the Ps
radius) depends on the asymmetry of the center-of-mass motion through $J$.
A simple perturbative estimate of the $J$ splitting of these states is
provided in Appendix~\ref{sec:pert}.

To define a spherically averaged collisional radius of Ps($2p$) we take a
weighted average of these data $(K_J,\rho_{2p[1,1]J})$ with weights $2J+1$ for
each $R_c$. The corresponding values are shown by diamonds in
Fig.~\ref{fig:Ps2p1+2+}. They lie close to the $J^\Pi=2^+$ date points
(see Appendix~\ref{sec:pert} for an analytical explanation),
and agree very well with the momentum dependence predicted by the $J^\Pi=1^-$
data. Using these together gives the linear fit
\begin{equation}\label{eq:rho2p}
\rho_{2p}(K) = 5.35 - 2.77K.
\end{equation}
The static radius for the $2p$ state $\rho_{2p}(0)=5.35$~a.u. is again close to
the mean radius of free Ps($2p$), i.e., $\frac12 \langle r\rangle =5$~a.u.

Regarding $J^\Pi=1^-$ data, the Ps($2p$) radii in the states with $L=0$ (two for
each of $R_c$) are naturally spherically averaged. The $J^\Pi=1^-$ states
with $L=2$ are parts of the $J$-dependent manifold ($J=1$, 2, and 3). Here
it appears that the $J=1$ state here is close to the $J$-averaged value
(see Appendix~\ref{sec:pert}), so that all $J^\Pi=1^-$ data follow the same
linear momentum dependence.


\subsection{Estimate of the cavity shift of Lyman-$\alpha $ transition}

Measuments of the Ps Lyman-$\alpha $ transition in porous silica revealed a
blue shift of the transition energy
$\Delta E=1.26\pm 0.06$~meV~\cite{cassidylymanalpha}. The pore size in this
material is estimated to be $a\sim 5~\text{nm}$~\cite{crivelly}. Assuming
spherical pores for simplicty, we find their radius $R_c\sim 50$~a.u. The Ps
center-of-mass momentum in the lowest energy state in such pores is
$K\simeq \pi /R_c\sim 0.06$~a.u. For such a small momentum one can use
static values of the Ps radius in $1s$ and $2p$ states (see Figs.~\ref{fig:Ps1s}
and \ref{fig:Ps2p1+2+}). Considering $S$-wave Ps ($L=0$), we estimate the
cavity shift of the Lyman-$\alpha $ transition energy from
Eq.~(\ref{eq:EnlNLrho}),
\begin{equation}\label{eq:cavshift}
\Delta E\simeq \frac{\pi ^2}{2R_c^3}(\rho _{2p}-\rho _{1s}).
\end{equation}
For static radii $\rho_{1s}=1.65$~a.u. and $\rho _{2p}=5.35$~a.u., and
$R_c=50$~a.u., we obtain $\Delta E=4$~meV. This value is close to the naive
estimate using mean Ps radii~\cite{cassidylymanalpha} and is significantly
larger than the experimental value.

It appears from the measured $\Delta E$ that the radius of Ps($2p$) is
only slightly greater than that of Ps($1s$). This effect is likely due to
the nature of Ps interaction with the wall in a real material. The Ps($2p$)
state is degenerate with Ps($2s$), and their linear combination (a hydrogenic
eigenstate in parabolic coordinates~\cite{landauQM}) can possess a permanent
dipole moment. Such a state can have a stronger, more attractive interaction
with the cavity wall than the ground state Ps($1s$). This interaction will
result in an additional phase shift of the Ps center-of-mass wave function
reflected by the wall. The scattering phase shift $\delta _{nl}(K)$ is related
to the Ps radius $\rho _{nl}(K)$ by $\delta _{nl}(K)=-K\rho _{nl}(K)$
[cf.~Eq.~(\ref{eq:def_rho})], with the static radius $\rho _{nl}(0)$ playing the
role of the scattering length. It is known that the van der Waals interaction
between the ground-state Ps and noble-gas atoms can significantly reduce the
magnitude of the scatetring length
\cite{mitroyivanov,mitroybromley03,fabrikant,psscattering}. It can be expected
that a similar dispersive interaction between excited-state Ps and the
cavity wall can reduce the effective radius of Ps($2p$) by more than that
of Ps($1s$), to produce the difference $\rho _{2p}-\rho_{1s}\approx 1$~a.u.
compatible with experiment.

\section{Conclusions}\label{sec:conclusion}

A $B$-spline basis was employed to obtain single-particle electron and
positron states within an otherwise empty spherical cavity. These states were
used to construct the two-particle states of positronium, including only
finitely many partial waves and radial states in the expansion. Diagonalization
of the Hamiltonian matrix allowed us to determine the energy and expectation
values of the electron-positron separation and contact density for each state.
Extrapolation of the energy with respect to the maximum orbital angular
momentum $\lmax $ and the number of radial states $\nmax $ included for each
partial wave was carried out. The electron-positron separation and contact
density values were
also extrapolated with respect to the number of partial waves included and used
to determine the nature (i.e., the quantum numbers) of each positronium state.
From the extrapolated energies, the effective collisional radius of the
positronium atom was determined for each state.

We have found that the radius of the Ps atom in the ground state has a linear
dependence on the Ps center-of-mass momentum, Eq.~(\ref{eq:rho1s}), the radius
being smaller for higher impact momenta. The radius of Ps($2s$) also displays
a linear momentum dependence, Eq.~(\ref{eq:rho2s}). The static (i.e.,
zero-momentum) collisional radii of the $1s$ and $2s$ states are 1.65 and
7~a.u., respectively.
Determining the effective radius of Ps in $2p$ state is more complex due to its
asymmetry. Spherically averaged values of the collisional radius are obtained
directly for Ps $S$-wave states in the cavity, with Ps $D$-wave states giving
similar radii. However, determining the corresponding values for the Ps
$P$-wave states required averaging over the total angular momentum of the
Ps states in the cavity. (See Appendix~\ref{sec:pert} for a quantitative 
explanation for the $J$ dependence of the $2p[1,1]$ and $2p[1,2]$ energy
levels.) After this, all data points were found to follow
the linear dependence on the Ps momentum, giving the static radius of 5.35~a.u.
In all three cases, the static values of the effective Ps radius are close to
the expectation value of the radius of free Ps, i.e., a half of the mean
electron-positron separation.

While the linear fits obtained here for the dependence of the effective Ps
radius on the center-of-mass momentum are clearly very good, particularly for
the $1s$ state, it must be noted that there is a certain amount of scatter
around the lines. This phenomenon may be due to numerical errors in the
two-particle-state calculations or in the extrapolation of the energy
eigenvalues (or both). The main issue here is the slow convergence of the
single-center expansion for states that describe the compact Ps atom away
from the origin. This issue also prevented us from perfoming calculations for
larger-sized cavities, which would provide effective Ps radii for lower
center-of-mass momenta. A possible means to reduce the scatter in the data
and tackle large cavities could be to include more partial waves and radial
states per partial wave in the CI expansion. However, with the Hamiltonian
matrix dimensions increasing as $\lmax \nmax ^2$, this quickly becomes
computationally expensive. An alternative would be to use a variational
approach with explicitly correlated two-particle wave functions.

Although we have only considered the $1s$, $2s$, and $2p$ states in the present
work, it is possible to use the method to investigate the effective radii of
higher excited states, e.g., the $3s$, $3p$ and $3d$ states. However, this
would require calculations with much larger cavities that can fit the
$n=3$ Ps states without significantly squeezing them.

It was noted earlier that confinement can cause a Ps atom to ``shrink'' from
its size \textit{in vacuo}. This manifests in the form of a reduced
electron-positron separation and increased contact density; these effects were
observed for the $2s$ and $2p$ states. While this is true in an idealized
hard-wall cavity, in physical cavities (e.g., in a polymer) there is a second,
competing effect. Polarization of the Ps atom by the
surrounding matter may actually lead to a swollen Ps atom, which causes the
contact density to be reduced from its value \textit{in vacuo}
\cite{dupasquier,mcmullen}. Experimentally, it has generally been found that
the net result of these two effects is that the contact density is
reduced from its \textit{in-vacuo} value, although increased values are
not necessarily impossible~\cite{mcmullen,brandt}. It may be possible to
determine more physical effective Ps radii by using realistic electron- and
positron-wall potentials in place of the hard wall we have used here.
Such developement of the approach adopted in the present work should yield
much more reliable data for the distorted Ps states than crude model
calculations~\cite{Tanzi16}.

The technique outlined in this paper is eminently suitable for implementing a
bound-state approach to low-energy Ps-atom scattering. By calculating
single-particle electron and positron states in the field of an atom at the
center of the cavity (rather than the empty cavity) and constructing
two-particle Ps wave functions from these, a shifted set of energy levels may
be found. From these, the Ps-atom scattering phase shifts can be determined
[cf.~Eq.~(\ref{eqn:phaseshifts})] using the now known collisional radius of Ps.
We have carried out several preliminary calculations in this area for elastic
Ps($1s$)-Ar scattering at the static (Hartree-Fock) level and found a
scattering length of 2.85~a.u., in perfect agreement with an earlier fixed-core
stochastic variational method calculation in the static approximation by
Mitroy and Ivanov~\cite{mitroyivanov}. This provides evidence that the linear
fits presented here account for the finite size of the Ps atom in scattering
calculations very accurately. In our calculations we have also observed
fragmented Ps states at higher energies (this manifests as a larger-than-usual
electron-positron separation). These have been ignored in the present work,
but it may be possible to use them to obtain information about inelastic
scattering using this method.

It is hoped that the results presented here will be of use in future studies of
both confined positronium and positronium-atom scattering.

\begin{acknowledgments}
We are grateful to D. G. Green for helpful comments and suggestions.
The work of A.R.S. has been supported by the Department for the Economy,
Northern Ireland.
\end{acknowledgments}

\appendix

\section{Working expressions for the Hamiltonian matrix and expectation
values}\label{sec:app1}

Written in terms of the angular and radial parts of the electron and positron
basis states, the Ps wave function is
\begin{align}
\Psi_{J\Pi}(\vec{r}_e,\vec{r}_p)&=\frac{1}{r_er_p}
\sum _{\substack{\mu ,\nu\\ m_\mu,m_\nu}} C^{(J)}_{\mu \nu}
P_\mu (r_e)P_\nu (r_p)\nonumber \\
&\quad{}\times C^{JM}_{l_\mu m_\mu l_\nu m_\nu} Y_{l_\mu m_\mu }(\Omega _e)
Y_{l_\nu m_\nu }(\Omega _p),\label{eq:Psi_J}
\end{align}
where $C^{JM}_{l_\mu m_\mu l_\nu m_\nu}$ is the Clebsch-Gordan coefficient, and
the indices $\mu $ and $\nu $ enumerate the radial electron and positron 
basis states with various orbital angular momenta,
$\mu \equiv \eps _\mu l_\mu $ and $\nu \equiv \eps _\nu l_\nu $. Besides the
selection rules due to the Clebsch-Gordan coefficient, the summation
is restricted by parity, $(-1)^{l_\mu +l_\nu}=\Pi $, where $\Pi=1$ $(-1)$ for the
even (odd) states.

Integration over the angular variables in the Coulomb matrix
elements is performed analytically~\cite{Varshalovich}, and the Hamiltonian matrix elements
for the Ps states with the total angular momentum $J$ are given by
\begin{align}
H^{(J)}_{\mu '\nu',\mu \nu}&=(\eps _\mu +\eps _\nu )
\delta_{\mu \mu'} \delta _{\nu\nu'}+
V^{(J)}_{\mu '\nu',\mu \nu},
\end{align}
where
\begin{equation}\label{eq:VJ}
V^{(J)}_{\mu '\nu',\mu \nu}
=\sum_l (-1)^{J+l}\sj{J}{l_{\mu '}}{l_{\nu '}}{l}{l_\nu }{l_\mu}
\langle \nu '\mu '\|V_l\|\mu\nu \rangle,
\end{equation}
and the reduced Coulomb matrix element is
\begin{align}
\langle \nu'\mu'\|V_l\|\mu \nu \rangle &=\sqrt{[l_{\nu'}][l_{\mu '}]
[l_\mu ][l_\nu ]}
\tj{l_{\mu '}}{l}{l_\mu}{0}{0}{0}
\tj{l_{\nu '}}{l}{l_\nu}{0}{0}{0}\nonumber \\ \label{dir}
&\quad{}\times \int_{0}^{R_c}\!\!\int_{0}^{R_c}
P_{\nu '}(r_p)P_{\mu '}(r_e)\frac{r_{<}^{l}}{r_{>}^{l+1}} \nonumber\\
&\quad {}\times P_\mu (r_e)P_\nu (r_p) \, dr_e \, dr_p,
\end{align}
with $[l]\equiv 2l+1$, $r_<=\min(r_e,r_p)$, and $r_>=\max(r_e,r_p)$.

The expectation value of the electron-positron separation 
$r_{ep}=\lvert\vec{r}_e-\vec{r}_p\rvert$ for an eigenstate
with eigenvector $C^{(J)}_{\mu \nu}$ is found as
\begin{align}\label{eq:r12}
\langle r_{ep}\rangle =\sum_{\substack{\mu',\nu'\\ \mu,\nu,l}} 
C^{(J)}_{\mu' \nu '}C^{(J)}_{\mu \nu}(-1)^{J+l}
\sj{J}{l_{\mu '}}{l_{\nu '}}{l}{l_\nu }{l_\mu}
\langle \nu'\mu' \Vert S_l \Vert \mu\nu\rangle ,
\end{align}
where
\begin{align}\label{eq:Sl}
\langle \nu'\mu' \Vert S_l \Vert \mu\nu\rangle &=
\sqrt{[l_{\nu'}][l_{\mu'}][l_\mu][l_\nu]}
\tj{l_{\mu '}}{l}{l_\mu}{0}{0}{0}
\tj{l_{\nu '}}{l}{l_\nu}{0}{0}{0}\nonumber \\
&\quad{}\times \int_0^{R_c} \!\!\int_0^{R_c}
P_{\nu'}(r_p) P_{\mu'}(r_e)
 \frac{r_<^l}{r_>^{l+1}} \nonumber \\
&\quad{}\times \left( \frac{r_<^2}{2l+3} - \frac{r_>^2}{2l-1}\right)
P_\mu(r_e) P_\nu(r_p) \, dr_e \, dr_p .
\end{align}

Similarly, the expectation value of the electron-positron contact density 
$\delta _{ep}=\delta(\vec{r}_e-\vec{r}_p)$ is 
\begin{align}\label{eq:del12}
\langle\delta_{ep}\rangle =\sum_{\substack{\mu',\nu'\\ \mu,\nu,l}} 
C^{(J)}_{\mu' \nu '}C^{(J)}_{\mu \nu}(-1)^{J+l}
\sj{J}{l_{\mu '}}{l_{\nu '}}{l}{l_\nu }{l_\mu}
\langle \nu' \mu' \| \delta_l \| \mu \nu \rangle ,
\end{align}
where
\begin{align}
\langle \nu' \mu' \Vert \delta_l \Vert \mu \nu \rangle &= 
\sqrt{[l_{\nu'}][l_{\mu'}][l_\mu][l_\nu]}
\tj{l_{\mu '}}{l}{l_\mu}{0}{0}{0}
\tj{l_{\nu '}}{l}{l_\nu}{0}{0}{0}\nonumber \\
&\quad{}\times \frac{[l]}{4\pi}\int_0^{R_c}P_{\nu'}(r)P_{\mu'}(r)P_\mu(r)P_\nu(r)
\frac{dr}{r^2} .
\end{align}

\section{Splitting of Ps $nl[N,L]$ states due to interaction with the cavity
wall}\label{sec:pert}

The angular part of the Ps wave function in the cavity,
Eq.~(\ref{eq:Psi_nlNL}), is
\begin{equation}
\Theta_{lL}^{(J)}(\Omega_\vec{r},\Omega_\vec{R}) = \sum_{m,M_L} C_{lmLM_L}^{JM}
Y_{lm}(\Omega_\vec{r}) Y_{LM_L}(\Omega_\vec{R}).
\end{equation}
The electron and positron repulsion from the wall is strongest when the vectors
$\vec{r}$ and $\vec{R}$ are parallel or antiparallel. In the simplest
approximation, we can take the corresponding perturbation as being proportional
to $\cos^2\theta $, where $\theta $ is the angle between
$\vec{r}$ and $\vec{R}$. Shifting this by a constant to make the spherical
average of the perturbation zero, we write it as
\begin{equation}\label{eq:deltaV}
\delta V(\Omega_\vec{r},\Omega_\vec{R})=\alpha P_2(\cos\theta ),
\end{equation}
where $\alpha$ is a constant that can depend on the quantum numbers $n$
and $N$ and on the cavity radius $R_c$, and $P_2$ is the second Legendre
polynomial. The corresponding energy shift is
\begin{equation}
\Delta E_J^\text{(pert)} = \alpha \iint \left\lvert
\Theta_{lL}^{(J)}(\Omega_\vec{r},\Omega_\vec{R})\right\rvert^2 P_2(\cos\theta) \,
d\Omega_\vec{r} \, d\Omega_\vec{R},
\end{equation}
by first-order perturbation theory. Intergating over the angles, one obtains
\cite{Varshalovich}
\begin{align}
\Delta E_J^\text{(pert)}&= \alpha (2l+1)(2L+1)
\tj l2l000
\tj L2L000
\nonumber\\
&\quad{}\times(-1)^J 
\sj LlJlL2 
\nonumber\\
 &= \alpha (2l+1)(2L+1) 
\tj l2l000
\tj L2L000
\nonumber\\
&\quad{}\times(-1)^{l+L} \sqrt{\frac{(2L-2)! \, (2l-2)!}{(2L+3)! \, (2l+3)!}}\,
\left( 6X^2 + 6X - 8Y\right),\label{eq:pert}
\end{align}
where $X= J(J+1)-l(l+1)-L(L+1)$ and $Y=l(l+1)L(L+1)$.

It is easy to check that the average energy shift is zero, i.e.,
\begin{equation}
\sum _J(2J+1)\Delta E_J^\text{(pert)}=0,
\end{equation}
as it should be for a perturbation with a zero spherical
average, $\langle \delta V\rangle =0$.

For Ps states $2p[1,1]$ and $2p[1,2]$ the possible values of $J$ are 0, 1, 2
and 1, 2, 3 respectively. In each case, let $E_J$ denote the calculated energy
eigenvalues of the $J$ manifold, with the average energy
\begin{equation}
\langle E_J \rangle = \frac{\sum_J (2J+1) E_J}{\sum_J (2J+1)}.
\end{equation}
To compare the numerical energy shifts
$\Delta E_J\equiv E_J-\langle E_J \rangle $ with $\Delta E_J^\text{(pert)}$,
we choose $\alpha $ to reproduce the calculated mean-squared shift, viz.,
\begin{equation}
\sum_J (2J+1) \left[\Delta E_J^\text{(pert)}\right]^2=\sum _J(2J+1)
\left[\Delta E_J\right]^2.
\end{equation}

Table~\ref{tab:deltaE} shows the energy eigenvalues, $J$-averaged energies,
numerical and perturbative energy shifts $\Delta E_J$ and
$\Delta E_J^\text{(pert)}$, as well as the values of $\alpha $, for
Ps($2p$) states with $L=1$ and $L=2$, for $R_c=10$ and 12~a.u. 
\begin{table*}
\caption{\label{tab:deltaE}Comparison of the energy shifts $\Delta E_J$
obtained from the numerical eigenvalues $E_J$ with the perturbative
estimates $\Delta E_J^\text{(pert)}$, Eq.~(\ref{eq:pert}), for Ps($2p$) states
with $N=1$ and $L=1,\,2$, for cavity radii $R_c=10$ and 12~a.u.}
\begin{ruledtabular}
\begin{tabular}{cccd{1.8}d{1.7}d{2.7}cd{1.6}d{2.6}}
$nl[N,L]$ & $R_c$ & $J^\Pi$ & \multicolumn{1}{c}{$E_J$} & \multicolumn{1}{c}{$\langle E_J \rangle$} & \multicolumn{1}{c}{$\Delta E_J$} & $\Delta E_J^{\text{(pert)}} / \alpha$ & \multicolumn{1}{c}{$\alpha$} & \multicolumn{1}{c}{$\Delta E_J^{\text{(pert)}}$}  \\
\hline
$2p[1,1]$ & 10 & $0^+$ & 0.0768631 &  & 0.0203912 & $2/5$ &  & 0.019022  \\
&& $1^+$ & 0.0477391 &0.0564719 &-0.0087328& $-1/5$ &0.047556& -0.009511 \\
&& $2^+$ & 0.0576334 & & 0.0011615 &$1/25$ && 0.001902  \\
\cline{2-9}
&12 & $0^+$ & 0.0253903 &  & 0.0126834 & $2/5$ &  & 0.011915  \\
&& $1^+$ & 0.00718021 &0.0127069 & -0.0055267 & $-1/5$ &0.029787& -0.005957  \\
&& $2^+$ & 0.0134862 & & 0.0007793 & $1/25$ && 0.001191  \\
\hline
$2p[1,2]$ & 10 & $1^-$ & 0.111199 &  & -0.000037 & $1/5$ &  & 0.008045  \\
&& $2^-$ & 0.103201 &0.111236 & -0.008035 & $-1/5$ &0.040224& -0.008045  \\
&& $3^-$ & 0.116992 & & 0.005756 & $2/35$ && 0.002299  \\
\cline{2-9}
& 12 & $1^-$ & 0.0507503 &  & 0.0008605 &$1/5$ &  & 0.005357  \\
&& $2^-$ & 0.0443476 & 0.0498898& -0.0055422 &$-1/5$ &0.026787& -0.005357  \\
&& $3^-$ & 0.0534797 & & 0.0035899 &$2/35$ && 0.001531  
\end{tabular}
\end{ruledtabular}
\end{table*}
To complete
the multiplet for $L=2$, calculations for $J^\Pi=2^-$ and $3^-$ were
carried out using $\lmax=9$--13 and $\nmax = 8$--13. For a given $R_c$ the values of $\alpha $ for $L=1$ and 2 states are similar. On the other hand, when $R_c$ increases from 10 to 12~a.u., the values of $\alpha $ decrease
as $1/R_c^q$ with $q\sim 2.5$. This is close to the expected $1/R_c^3$ dependence of the energy shifts with the cavity radius~\cite{greenreply}.

\begin{figure}[tb]
\includegraphics[width=\columnwidth]{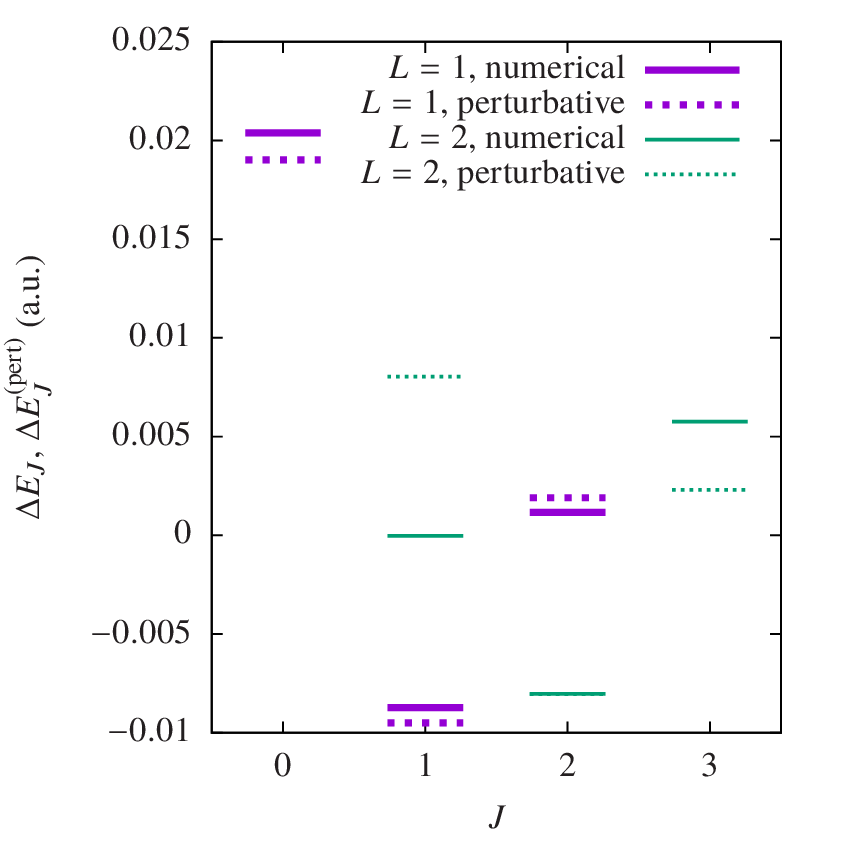}
\caption{\label{fig:J_split_R10}Values of $\Delta E_J$ and $\Delta E_J^\text{(pert)}$ for the $2p[1,1]$ and $2p[1,2]$ states in a cavity of radius $R_c=10$~a.u.}
\end{figure}

\begin{figure}[tb]
\includegraphics[width=\columnwidth]{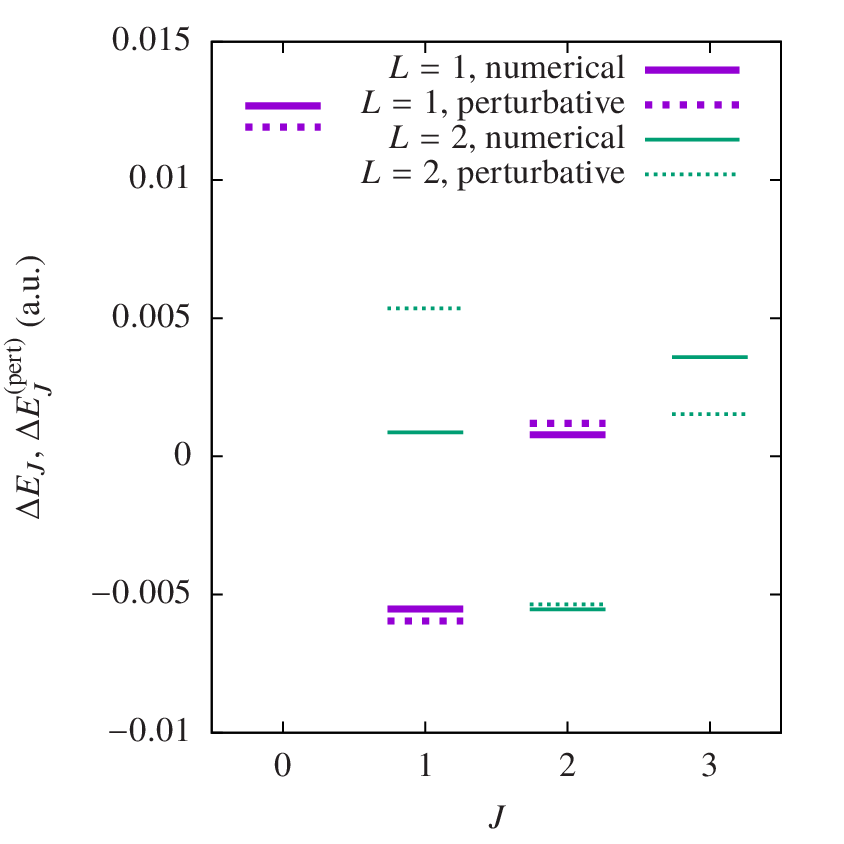}
\caption{\label{fig:J_split_R12}Values of $\Delta E_J$ and $\Delta E_J^\text{(pert)}$ for the $2p[1,1]$ and $2p[1,2]$ states in a cavity of radius $R_c=12$~a.u.}
\end{figure}

Figures~\ref{fig:J_split_R10} and \ref{fig:J_split_R12} compare the values of $\Delta E_J$ with their perturbative estimates $\Delta E_J^\text{(pert)}$ for $R_c=10$~a.u. and 12~a.u. respectively. 
For $L=1$ the perturbative estimates of the energy shifts are in excellent agreement with their numerical counterparts. Note that the small shift of the $J^\Pi=2^+$ level is explained by the small magnitude of the corresponding $6j$ symbol in Eq.~(\ref{eq:pert}). For $L=2$ states, the perturbative estimate reproduces the overall $J$ dependence of the calculated energy shift, with the $J=2$ state being the lowest of the three. However, the relative positions of the $J^\Pi=1$ and 3 states are reversed. This is probably due to higher-order corrections or level mixing not described by Eq.~(\ref{eq:deltaV}).
Note that the numerical shift is smallest for the $J=1$ state, which justifies its use in determining the fit (\ref{eq:rho2p}).


%

\end{document}